\documentclass[prl,amsmath,amssymb,twocolumn,superscriptaddress]{revtex4-2}

\usepackage{graphicx}
\usepackage{times}
\usepackage{float}
\usepackage{color}
\graphicspath{{main_figures},{suppl_figures},{Additional_figures}}

\begin{document}

\newcommand{\red}[1]{\color{red}{#1}}

\title{Confined and deconfined spinon excitations in the rectangular-lattice  
quantum antiferromagnet}

\author{N. E. Shaik}
\affiliation{Laboratory for Quantum Magnetism, Institute of Physics, 
Ecole Polytechnique F\'ed\'erale de Lausanne (EPFL), CH-1015 Lausanne, 
Switzerland}

\author{E. Fogh}
\affiliation{Laboratory for Quantum Magnetism, Institute of Physics, 
Ecole Polytechnique F\'ed\'erale de Lausanne (EPFL), CH-1015 Lausanne, 
Switzerland}

\author{B. Dalla Piazza}
\affiliation{Laboratory for Quantum Magnetism, Institute of Physics, 
Ecole Polytechnique F\'ed\'erale de Lausanne (EPFL), CH-1015 Lausanne, 
Switzerland}

\author{B. Normand}
\affiliation{Laboratory for Quantum Magnetism, Institute of Physics, 
Ecole Polytechnique F\'ed\'erale de Lausanne (EPFL), CH-1015 Lausanne, 
Switzerland}
\affiliation{PSI Center for Scientific Computing, Theory and Data, CH-5232 
Villigen-PSI, Switzerland}

\author{D. A. Ivanov}
\affiliation{Institute for Theoretical Physics, ETH Z\"urich, CH-8093 
Z\"urich, Switzerland }

\author{H. M. R{\o}nnow}
\affiliation{Laboratory for Quantum Magnetism, Institute of Physics, 
Ecole Polytechnique F\'ed\'erale de Lausanne (EPFL), CH-1015 Lausanne, 
Switzerland}


\begin{abstract}
Fractionalization remains one of the most fascinating manifestations of 
strong interactions in quantum many-body systems. In quantum magnetism, 
the existence of spinons -- collective magnetic excitations that behave as 
quasiparticles with fractional quantum numbers -- is proven in spin chains, 
but the criteria for their appearance in higher dimensions remain disputed. 
Motivated by experiments reporting the observation of spinons at high energies 
in the square-lattice Heisenberg antiferromagnet, we adopt the approach of 
extrapolating from where spinons are well defined. We study the dynamical 
properties of a Gutzwiller-projected wave function, the staggered-flux state, 
on a rectangular spin-1/2 Heisenberg lattice as a function of the spatial 
coupling ratio, $\gamma = J_y/J_x$. By studying the spectrum and the 
spinon separation distribution we show how, as the system evolves from 
one-dimensional (1D) towards 2D, the spinons become progressively more 
confined over most of reciprocal space, but remain deconfined at specific 
wave vectors. 
\end{abstract}

\maketitle

\begin{figure*}[t]
\includegraphics[width=0.96\textwidth]{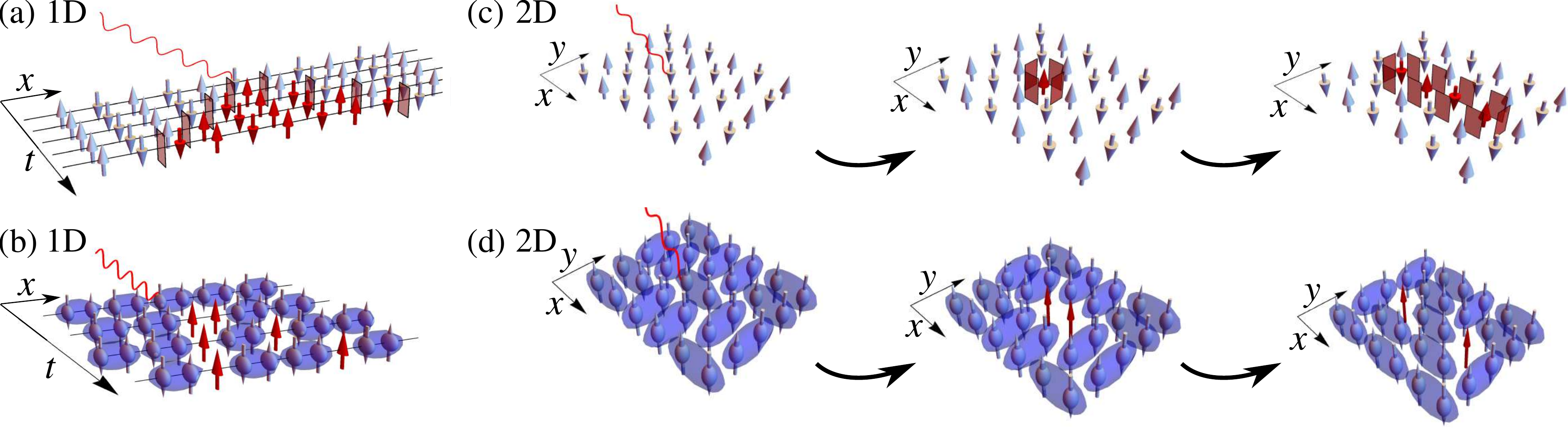}
\caption{{\bf Schematic representations of the propagation of spin-flip 
excitations in low-dimensional quantum spin systems.} (a) A spin flip in a 
1D N\'eel state creates two domain walls, each of which can be interpreted 
as a spinon. (b) A spin flip in the nearest-neighbor valence-bond state on 
a chain creates one dimer triplet, whose two uncorrelated spins propagate 
at no additional energy cost on the singlet background. (c,d) Analogous 
representations on the AFM square lattice. The propagation of two domain 
walls in a N\'eel state creates two lines of ferromagnetic bonds, leading 
to localization (c). By contrast, in a valence-bond state (d) the two freed 
spins can still propagate at no extra cost.} 
\label{fig:fig1}
\end{figure*}

\textit{Introduction.--}The emergence of quasiparticles with fractional 
quantum numbers is a fascinating collective phenomenon. Notable examples 
include quarks confined as hadrons \cite{quark_conf}, fractionally 
charged quasiparticles in the quantum Hall effect \cite{Laughlin}, 
spin-charge separation of electrons in one-dimensional (1D) systems 
\cite{Lieb}, solitons in polyacetylene-type molecules \cite{Su}, and 
spinons in antiferromagnetic (AFM) spin-1/2 Heisenberg chains 
\cite{Tennant,Lake2005,Mourigal}. While the fractionalization of spin 
excitations in 1D, represented schematically in Figs.~\ref{fig:fig1}(a-b),
is clearly demonstrable in theoretical models, the existence of true 
spinons in higher dimensions remains a challenge. Ideal fractionalization, 
meaning on all energy scales, is present in certain 2D models that have 
yet to be realized in quantum materials, but in known systems is replaced 
by conventional behavior as the two $S = 1/2$ fractions become confined 
into $S = 1$ magnons. While the search continues for materials displaying 
complete deconfinement as a consequence of sufficiently strong quantum 
fluctuations, for example when frustration effects act to suppress 
competing magnetic order, a less strict criterion for fractionalization 
is the observation of spinonic fingerprints at finite energies.

The idea that magnetic correlations in 2D systems can be dominated by the 
quantum fluctuations clearly manifest in 1D was proposed originally in the 
context of high-temperature superconductivity (HTS) \cite{AndersonSci}. 
The associated concepts, illustrated in Fig.~\ref{fig:fig1}(c), provided a 
starting point for the study of quantum spin liquids \cite{Savary2017,
Ng2017}. Although the ground state of the quantum ($S = 1/2$) square-lattice 
Heisenberg AFM relevant to HTS has long-ranged N\'eel order and sharp magnon 
bands at low energies, which have been verified quantitatively in a number 
of materials \cite{Keimer1992,Ronnow,Kim2001,Lancaster2007}, detailed 
spectroscopic measurements \cite{Coldea2001,Ronnow,Lumsden2006,Tsyrulin2009,
Headings2010,Christensen,Plumb2014,DallaPiazza} have found anomalous, 
continuum-like scattering at higher energies. Multiple theoretical studies 
have debated whether these results indicate extensive magnon bound-state 
formation \cite{Singh1995,Sandvik2001,Zheng2005,Powalski,Powalski2018} or 
a possible deconfinement of single magnons into fractional constituents 
\cite{AA1988,Hsu1990,Ho2001,DallaPiazza,Shao2017,Zhang2022}. 

In this Letter we perform a systematic analysis of fractionalization in the 
excitation spectrum of the $S = 1/2$ Heisenberg model on a rectangular lattice, 
i.e.~with no frustration. To understand if and how the reports of deconfinement 
at $(\pi,0)$ in the 2D limit are connected to complete deconfinement in 1D, we 
build on prior work establishing the validity of the staggered-flux state as 
an accurate description of the ground state of the rectangular-lattice model 
at all values of the spatial coupling ratio, $J_y/J_x$. We analyze 
the excitation spectrum of the staggered-flux state by using a Monte Carlo 
method to compute the transverse dynamical structure factor and spinon pair 
separation. Starting from a quantitatively accurate description of the 1D 
limit, we demonstrate that increasing $\gamma$ causes a systematic spreading 
of confinement in reciprocal space, but explicitly not in energy, until in the 
square lattice the deconfined nature remains only at the ($\pi,0$) points. 

We first review the concept of spin fractionalization with the schematic 
images of Fig.~\ref{fig:fig1}. When a single spin is flipped in a 
1D N\'eel state, it can be regarded as two domain walls whose separation costs 
no further energy [Fig.~\ref{fig:fig1}(a)]. If the macroscopic singlet ground 
state of a Heisenberg spin chain is visualized as a superposition of 
short-ranged dimer singlets [only nearest-neighbor valence bonds are shown 
in Fig.~\ref{fig:fig1}(b)], a single spin excitation creates a dimer triplet 
whose two constituent spins are again free to delocalize. These $S = 1/2$ 
excitations were established theoretically by the Bethe Ansatz \cite{Bethe} 
and have been observed in experiment \cite{Tennant,Lake2005,Mourigal}. However,
extending the domain-wall image to a 2D N\'eel state [Fig.~\ref{fig:fig1}(c)] 
shows clearly why a spin flip should preserve $S = 1$ (magnonic) character, 
with the 1D string of FM bonds creating linear confinement of putative spinons.
This has led to the notion that geometrical frustration strong enough to 
create a disordered (quantum spin-liquid \cite{Savary2017}) ground state 
[Fig.~\ref{fig:fig1}(c)] is required to allow fractionalization in 2D.

From all the reports of zone-boundary scattering anomalies in $S = 1/2$
square-lattice materials \cite{Coldea2001,Ronnow,Lumsden2006,Tsyrulin2009,
Headings2010,Christensen,Plumb2014,DallaPiazza}, the system thought to best 
realize the minimal, nearest-neighbor-only AFM Heisenberg model is copper 
formate tetradeuterate (CFTD) \cite{Ronnow}. Detailed measurements have 
revealed \cite{Christensen,DallaPiazza} that the magnon branch of a 
renormalized spin-wave theory survives as a sharp mode everywhere within 
and on the boundary of the Brillouin zone, except at wave vector $(\pi,0)$, 
where it loses spectral weight in favor of an apparent scattering continuum 
at higher energies. The most detailed theoretical and numerical studies of 
the spectrum, whether based on magnon bound states \cite{Powalski,Powalski2018}
or on fractionalization \cite{DallaPiazza,Shao2017,Zhang2022}, have also 
highlighted the special nature of the $(\pi,0)$ points.

\textit{Staggered-flux state on a rectangular lattice.--}It was shown in 
Ref.~\cite{DallaPiazza} that the staggered-flux state \cite{Hsu1990,Ho2001} 
on the square lattice exhibits continuum-like scattering and excitations 
akin to deconfined spinons at the $(\pi,0)$ points \cite{DallaPiazzath}. To 
establish whether a spinon interpretation is viable, and if so to connect 
this putative finite-energy deconfinement in 2D with the complete 
fractionalization observed in 1D, we therefore adopt the staggered-flux 
framework. We use it to perform an explicit study of the spectrum of the 
Heisenberg model on the rectangular lattice (represented in the inset of 
Fig.~\ref{fig:fig4}), 
\begin{equation}
\mathcal{H} = \sum_{\bf i}
J_x {\bf S}_{\bf i} \! \cdot \! {\bf S}_{{\bf i} + {\hat{\bf x}}} +
J_y {\bf S}_{\bf i} \! \cdot \! {\bf S}_{{\bf i} + {\hat{\bf y}}},
\label{H-physical}
\end{equation}
by altering the coupling ratio, $\gamma = J_y/J_x$, to traverse smoothly 
from a quasi-1D to a 2D system \cite{Sakai1989,Azzouz1993,Parola1993,
Miyazaki}.

The staggered-flux state is obtained from the mean-field decoupling of an  
Ansatz describing the system in terms of fermionic spinons, as we summarize 
in Sec.~S1A of the Supplemental Materials (SM) \cite{sm}. The mean-field 
state, $\big| \psi_{\rm SF} \big\rangle$, allows double site occupancy and 
a Gutzwiller projection, $\big| \psi_{\rm GS} \big\rangle = P_{\rm G} \big| 
\psi_{\rm SF} \big\rangle$, is performed to obtain the ground state in the 
physical state space as a variational wave function. In Ref.~\cite{Shaik2020}, 
we optimized the staggered flux in order to discuss the ground state of this 
rectangular-lattice Heisenberg model [Eq.~(\ref{H-physical})] as a function 
of $\gamma$. Although the staggered-flux state cannot reproduce the exact 
ground state of the square-lattice model, most notably lacking an ordered 
moment, we will show that it offers a valuable starting point for constructing 
this state in the sense that its finite-energy features, including fractional 
excitations, provide an excellent account of the spectral function. 

\begin{figure*}[t]
\includegraphics[width=0.9\textwidth]{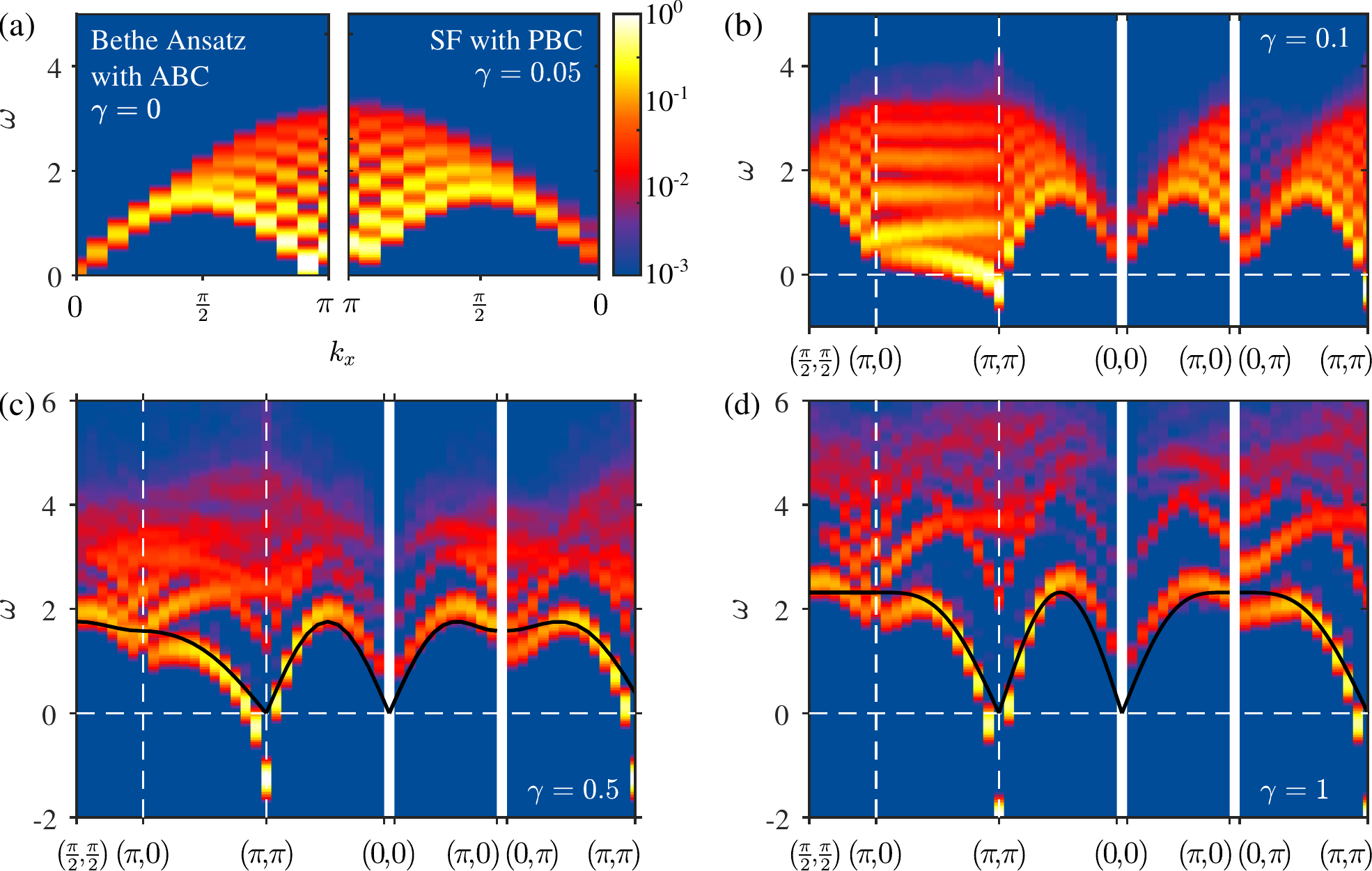}
\caption{{\bf Spectral functions on the rectangular lattice.} (a) Comparison
between the transverse dynamical structure factors (TDSFs) computed for a 
chain of $L = 24$ sites (i.e.~$\gamma = 0$) using the algebraic Bethe Ansatz 
(left) and the staggered-flux state with $\gamma = 0.05$ (right). (b-d) TDSFs, 
calculated with PBCs in $x$ and ABCs in $y$, on a selected ${\bf k}$-path for 
coupling ratios $\gamma = 0.1$ (b), $\gamma = 0.5$ (c), and $\gamma = 1$ (d) 
at system size $L = 24$. Black lines show the dispersion obtained from the 
self-consistent spin-wave theory described in Sec.~S3 of the SM \cite{sm}. 
The different BCs account for the asymmetry between $(\pi,0)$ and $(0,\pi)$ 
in panel (d).} 
\label{fig:fig2}
\end{figure*}

In the variational approach one approximates the physical excitations 
by projected fermionic excitations \cite{Hsu1990,Ho2001}. Following 
Ref.~\cite{DallaPiazza}, we construct transverse (spin-flip) excitations as 
linear combinations of fermionic particle-hole pairs,
\begin{equation}
\label{ekqbasis}
\big| \psi^n_{\bf q} \big\rangle = \sum_{\bf k} \phi_{\bf {kq}}^n | {\bf k}, 
{\bf q} \rangle, {\rm with \ } 
| {\bf k}, {\bf q} \rangle = P_{\rm G} \, d_{{\bf k}\uparrow}^{+\dagger} 
d^-_{{\bf k}-{\bf q}\downarrow} \big| \psi_{\rm SF} \big\rangle,
\end{equation}
where $d^-_{{\bf k} - {\bf q} \downarrow}$ destroys a down spin in the lower 
band, $d_{{\bf k} \uparrow}^{+\dag}$ creates an up spin in the upper band, 
and the coefficients $\phi^n_{\bf kq}$ are obtained by diagonalizing the 
Hamiltonian (\ref{H-physical}) projected onto the non-orthogonal states 
$| {\bf k},{\bf q} \rangle$; further details are provided in Sec.~S1B of 
the SM \cite{sm}. We adopt a Monte Carlo sampling technique to solve the 
associated generalized eigenvalue problem (Sec.~S1C of the SM \cite{sm})
and hence to compute the transverse dynamical structure factor (TDSF), which 
we approximate as 
\begin{equation}
\label{etdsf}
S^{\pm} ({\bf q},\omega) = \sum_n \big| \big\langle \psi_{\bf q}^n | S_{\bf 
q}^+ | \psi_{\rm GS} \big\rangle \big|^2 \delta (\omega - E^n_{\bf q} + 
E_{\rm GS}), 
\end{equation}
where $E_{\rm GS}$ and $E_{\bf q}^n$ are respectively the variational energies
of the ground and excited states. Thus we perform Monte Carlo simulations on 
systems of size $N = L$$\times$$L$, with $L$ up to 28, working in the space 
of all singly occupied states with $N/2 + 1$ up spins and $N/2 - 1$ down 
spins, and using a statistical average of the energies of the diagonalized 
Hamiltonian to obtain a suitably ${\bf k}$-resolved spectrum. Because the 
staggered-flux wave function is ill-defined at ${\bf k}$-points $(\pm \pi/2,
\pm \pi/2)$, we avoid these by keeping antiperiodic boundary conditions (ABCs) 
in $y$ while working with both ABCs periodic BCs (PBCs) in $x$, which is 
valuable both for comparison with the exact solution in 1D and for studying 
finite-size effects in our calculations. 

\begin{figure*}[t]
\includegraphics[width=0.88\textwidth]{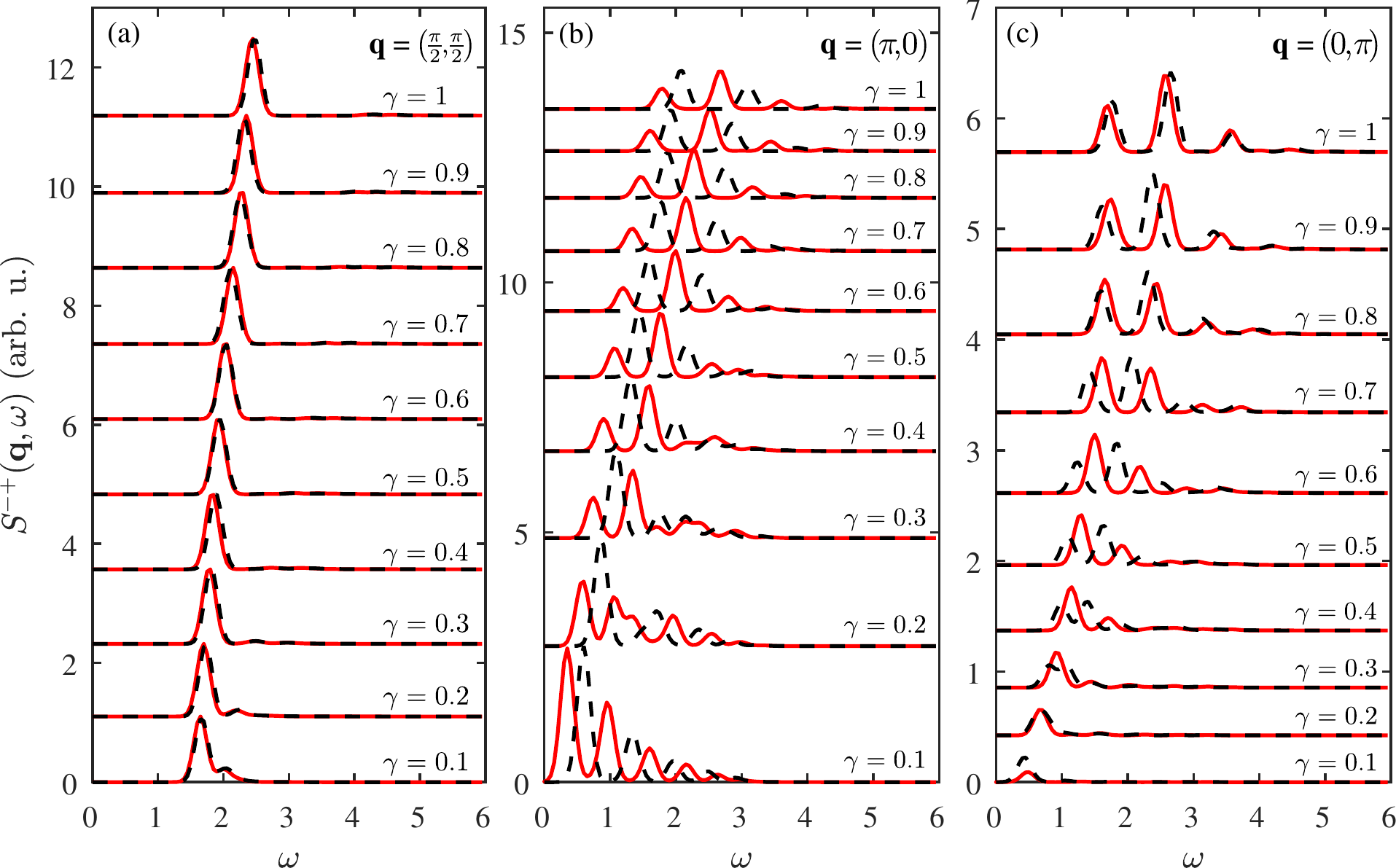}
\caption{{\bf Evolution of the TDSF with $\gamma$.} TDSF calculated with 
$L = 28$ at $(\pi/2,\pi/2)$ (a), $(\pi,0)$ (b), and $(0,\pi)$ (c). The black 
lines represent calculations performed with PBCs in $x$ and ABCs in $y$, 
the red lines ABCs in both directions.}
\label{fig:fig3}
\end{figure*}

To characterize the spatial profile of these excitations at each 
${\bf q}$, we introduce the weighted spatial spinon density,
\begin{equation}
\label{spinsepar}
\rho^q ({\bf r}) = \sum_n \frac{| \langle \psi^+_{\bf q} ({\bf r}) | \psi^n
_{\bf q} \rangle \langle \psi^n_{\bf q} | \psi_{\bf q} ({\bf 0}) \rangle 
|^2}{\langle \psi^+_{\bf q} ({\bf r}) | \psi^+_{\bf q} ({\bf r}) \rangle},
\end{equation}
where $| \psi^+_{\bf q} ({\bf r}) \rangle$ represents a pair of spinons 
induced at momentum transfer ${\bf q}$ with a separation ${\bf r} = (x,y)$ 
(Sec.~S1B of the SM \cite{sm}). As a means of distinguishing a possible 
spinon delocalization in 2D from the 1D delocalization we approach in the 
small-$\gamma$ limit, we define the quantities
\begin{equation}
\label{dmeas}
D_x ({\bf q}) = \sqrt{\sum_{\bf r} x^2 \rho^q ({\bf r})} \,\, {\rm and} 
\,\, D_y ({\bf q}) = \sqrt{\sum_{\bf r} y^2 \rho^q ({\bf r})}.
\end{equation}
These measure the RMS value of the spinon separation in one lattice direction 
and hence serve to gauge the degree and nature of spinon deconfinement. 

\textit{Results.--}Diagonalizing the projection of $\mathcal{H}$ onto the 
$| {\bf k}, {\bf q} \rangle $ basis yields $N$ discrete energies at each 
value of ${\bf q}$, which we convolve with Gaussians of width $\sigma = 0.1 
J_x$ to obtain the spectra we show in Fig.~\ref{fig:fig2}. We begin by 
benchmarking the validity of the staggered-flux state as a description of 
the spectrum in the 1D limit, where ideal (fully deconfined) spinons form 
the basis for the exact solution \cite{Bethe}. Figure \ref{fig:fig2}(a) 
compares the TDSF calculated using the algebraic Bethe Ansatz \cite{Caux_2005} 
with that of a staggered-flux state at $\gamma = 0.05$ (the lowest coupling we 
use) and $q_y = 0$. We find that the agreement is not only excellent, but 
quantitatively accurate in energy and intensity when an $L$-site chain is 
compared with an $L$$\times$$L$ rectangular lattice. The $\gamma = 0.05$ 
spectrum lies higher in energy by precisely the finite inter-chain coupling. 
The fermionic nature of the staggered-flux wave function makes the Bethe 
Ansatz solution with ABCs equivalent to the staggered-flux state with PBCs 
and vice versa \cite{Shaik2020}, which is reflected in the $2\pi/L$ shift in 
the position of the Goldstone mode near $k_x = \pi$ in Fig.~\ref{fig:fig2}(a).
This confirmation lays the foundation for a systematic study of spinon 
confinement effects when $\gamma$ is used to bring the system from 1D to 2D. 

Figures \ref{fig:fig2}(b-d) show the TDSF for the three coupling ratios 
$\gamma = 0.1$, 0.5, and 1 to capture the evolution of the spin dynamics 
with increasing two-dimensionality. The unphysical negative energies at 
$(\pi,\pi)$ reflect how the staggered-flux state, with no long-ranged order, 
fails to capture the Goldstone mode and the neighboring low-energy spectrum. 
At all other ${\bf k}$-points we observe how a quasi-continuum at small 
coupling ratios [Fig.~\ref{fig:fig2}(b)] evolves into sets of discrete 
modes [Fig.~\ref{fig:fig2}(d)]. The loss of spectral weight in the high-energy 
modes with increasing $\gamma$ occurs everywhere except at $(\pi,0)$, where 
these modes continue to provide almost half of the spectral weight at $\gamma
 = 1$. In Sec.~S2 of the SM \cite{sm} we present a finite-size comparison of 
different ${\bf k}$-points in this case. In Figs.~\ref{fig:fig2}(c,d) we show 
also the magnon branch obtained from a self-consistent spin-wave theory 
\cite{Shaik}, which provides an energetic renormalization factor, 
$Z_c({\bf k})$, to linear spin-wave theory as described in Sec.~S3 of the 
SM \cite{sm}; away from the negative-energy points we find a rather good 
match to the low-energy mode of the staggered-flux state.

Figure \ref{fig:fig3} shows the evolution of the TDSF calculated at $L = 28$ 
for the three high-energy points $(\pi/2,\pi/2)$, $(\pi,0)$, and $(0,\pi)$. 
At $(\pi/2,\pi/2)$ we observe the generic behavior of weak additional modes 
above the lowest one at low $\gamma$, which lose all their weight as the 
spectrum evolves to contain only a single magnon. At $(\pi,0)$, by contrast, 
the additional modes retain their weight even at $\gamma = 1$, and we show 
in Sec.~S2 of the SM \cite{sm} that the real band minimum (the magnon 
branch) is given by the PBC result in Fig.~\ref{fig:fig3}(b), verifying a 
conjecture made in Ref.~\cite{DallaPiazza}. Similar behavior is observed at 
$(0,\pi)$ [Fig.~\ref{fig:fig3}(c)], although here the spectral weight grows 
from zero at the 1D limit to fewer, stronger ``continuum'' modes due to our 
system size. Extending the insight of Fig.~\ref{fig:fig3} back to 
Figs.~\ref{fig:fig2}(b-d), our results suggest that in the thermodynamic 
limit the spectrum forms a single, sharp magnon-like mode as $\gamma 
\rightarrow 1$, at all ${\bf k}$-points other than $(\pi,0)$ and $(0,\pi)$; 
there we find that a series of discrete modes persists, whose number 
increases with $L$ (Sec.~S2 of the SM \cite{sm}), indicating the formation 
of a 1D-type continuum.

\begin{figure}[t]
\includegraphics[width=\columnwidth]{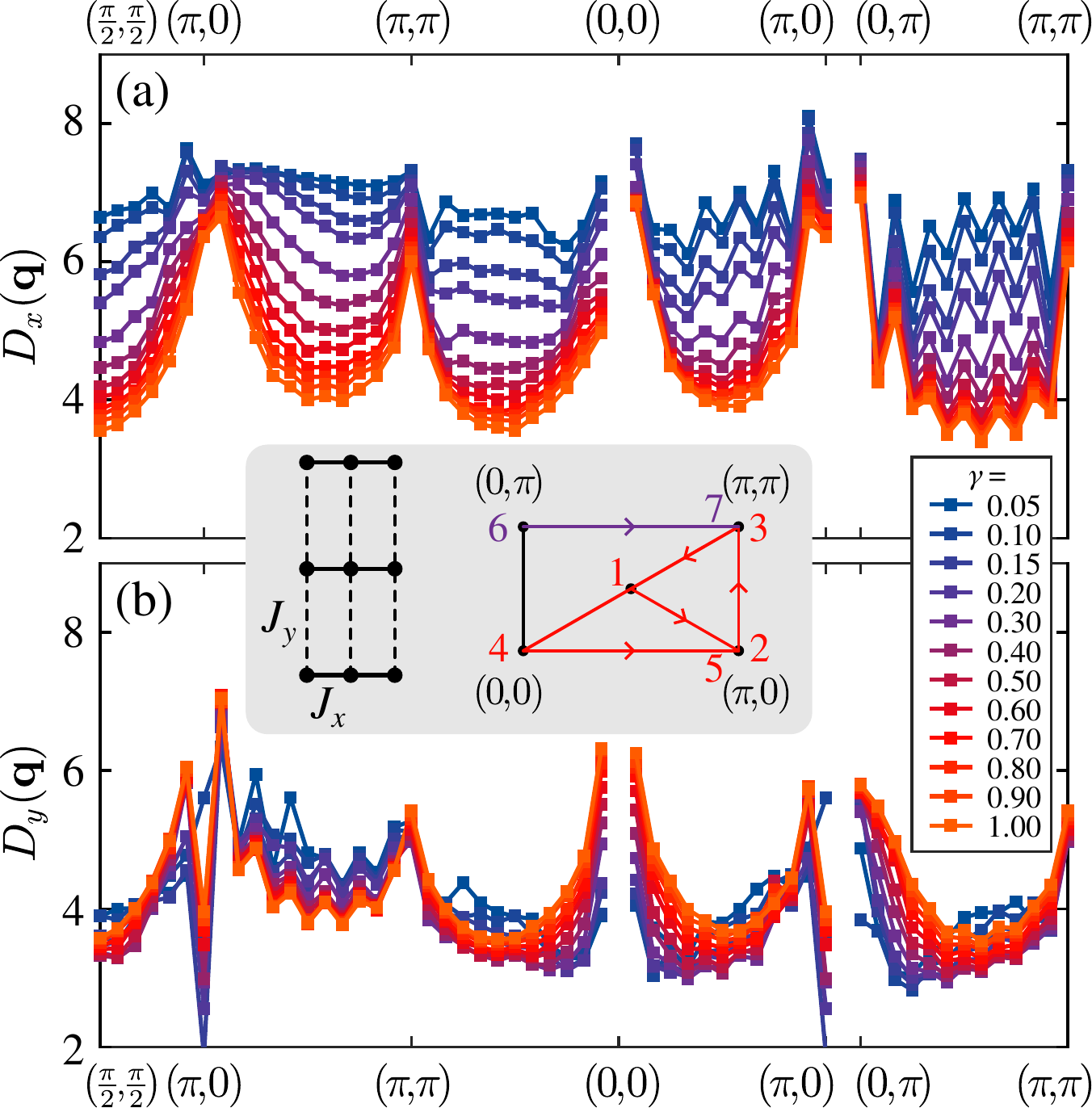}
\caption{{\bf RMS spinon pair separation.} Pair-separation parameters $D_x 
({\bf q})$ (a) and $D_y ({\bf q})$ (b) calculated on a system of $L = 24$ 
with PBCs in $x$, shown on the same ${\bf q}$-space path as in 
Fig.~\ref{fig:fig2} for different coupling ratios, $\gamma$. The inset shows 
a schematic representation of the rectangular lattice with superexchange couplings 
$J_x$ and $J_y$, and of the corresponding reciprocal space; numbers 1-7 indicate 
the order in which the ${\bf q}$-space path is followed.}
\label{fig:fig4}
\end{figure}

To interpret the line shapes in the TDSF, in Fig.~\ref{fig:fig4} we show the 
two RMS spinon separations $D_x ({\bf q})$ and $D_y ({\bf q})$ over the 
high-symmetry directions of ${\bf q}$-space for the full range of coupling 
ratios. As $\gamma$ is increased, $D_x ({\bf q})$ decreases at most ${\bf 
q}$-points, reflecting a progressive spinon confinement in the chain direction 
as a consequence of interchain coupling [Fig.~\ref{fig:fig4}(a)]. By contrast, 
$D_x ({\bf q})$ provides an excellent illustration of how $\gamma$ has 
no effect on confining spinon pairs at wave vectors $(\pi,0)$ and $(0,\pi)$. 
Concerning the analogous result at $(\pi,\pi)$ and $(0,0)$, these low energies 
are where the staggered-flux state without in-built AF order \cite{DallaPiazza}
captures neither the Goldstone mode nor, presumably, the confining effect of 
the ordered state on the spinons, and hence we do not consider these points 
further. For further insight into the meaning of deconfinement at $(0,\pi)$ 
in this geometry, Fig.~\ref{fig:fig4}(b) shows that $D_y ({\bf q})$ is almost 
invariant with $\gamma$, indicating how the trivial spinon localization caused 
by small interchain coupling turns into a real localization due to the 2D 
character of the system, with the exception of the growth in RMS separation 
around the $(0,\pi)$ and $(\pi,0)$ [and $(\pi,\pi)$ and $(0,0)$] points. 

\begin{figure}[t]
\includegraphics[width=\columnwidth]{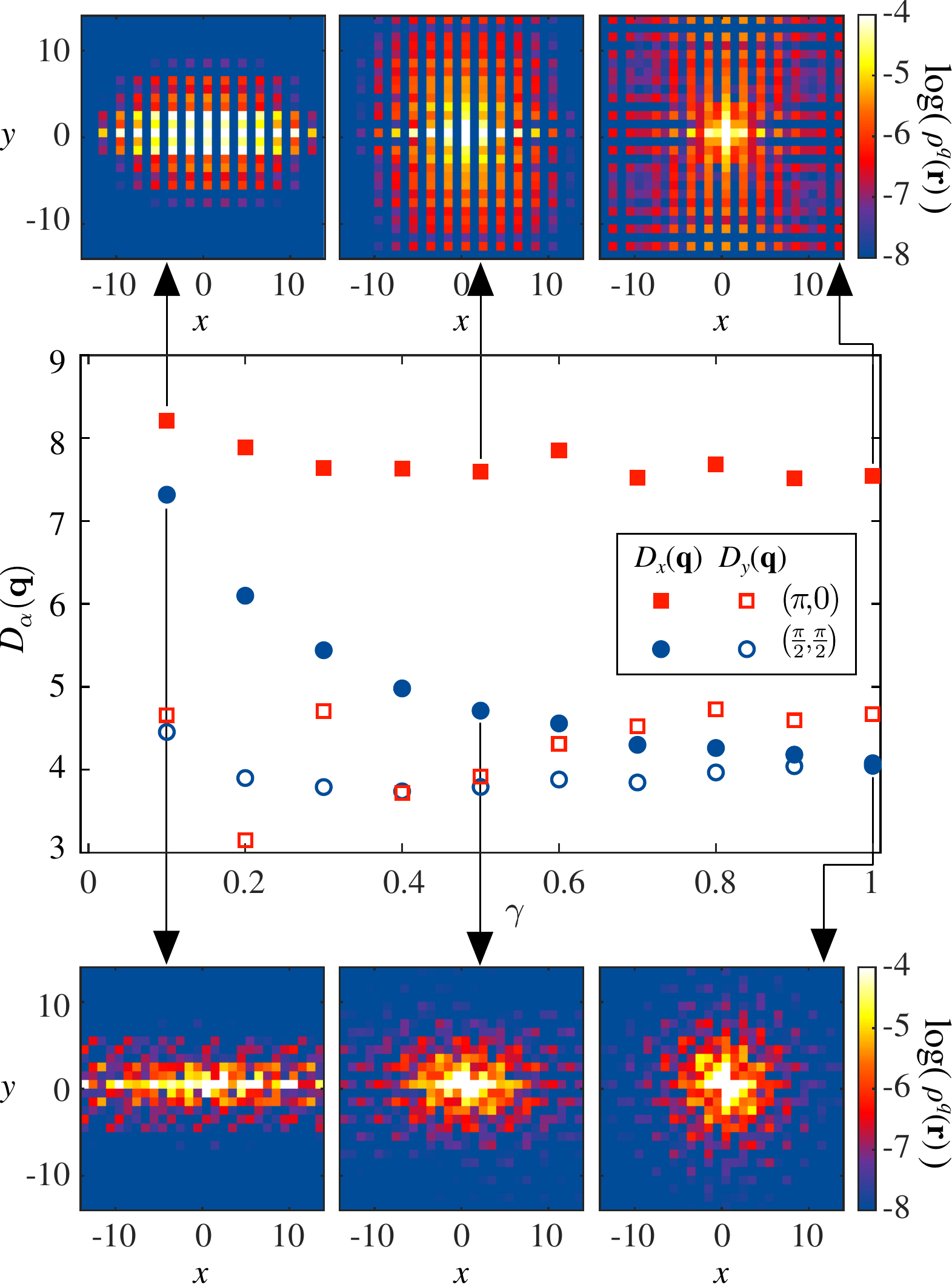}
\caption{{\bf Spinon pair separation on the rectangular lattice.} $D_x({\bf 
q})$ and $D_y({\bf q})$ calculated on a system of $L = 28$ with PBCs in 
$x$, shown as a function of $\gamma$ at ${\bf q} = (\pi,0)$ and ${\bf q}
 = (\pi/2,\pi/2)$. The subpanels show the corresponding spinon 
pair-separation distributions, $\rho^q ({\bf r})$.}
\label{fig:fig5}
\end{figure}

For a maximally clear visualization of the different behavior of the spinon 
pair separation, we focus on the points $(\pi/2,\pi/2)$ and $(\pi,0)$ to 
show the RMS separations as a function of $\gamma$ in Fig.~\ref{fig:fig5}. 
In the subsidiary panels we show the corresponding spinon pair density 
distributions at $\gamma = 0.1$, 0.5, and 1 on a 28$\times$28 lattice. At 
$(\pi/2,\pi/2)$, we observe a gradual decrease of $D_x ({\bf }q)$ and a clear 
localization of the pair density around $r = 0$ that has almost saturated 
at $\gamma = 0.5$. As with all ${\bf q}$-points other than the exceptions 
visible in Fig.~\ref{fig:fig4}, this is the characteristic sign of spinon 
confinement, even though the ground state (the staggered-flux state) has 
no magnetic order. Again the $(\pi,0)$ point is completely different, with 
$D_x ({\bf q})$ remaining flat and the system-scale 1D distribution turning 
into a system-scale 2D distribution, indicating that the spinons remain 
fully deconfined.

\textit{Conclusions.--}We have studied the crossover from 1D to 2D physics 
in the $S = 1/2$ Heisenberg model with only nearest-neighbor coupling and no 
frustration. We base our investigation on a Gutzwiller-projected variational 
wave function that yields a staggered-flux ground state, and thus we obtain 
results that are highly indicative but not definitive. By comparison with the 
Bethe Ansatz solution, we demonstrate that the staggered-flux state provides 
a quantitatively accurate description of the excitation spectrum in the 1D 
limit, reproducing the des Cloizeaux-Pearson-Faddeev spinon continuum. 
Increasing the interchain coupling to the 2D limit causes a systematic 
vanishing of the spectral weight in all the higher-energy modes at almost 
all ${\bf q}$-vectors, leaving only a well-defined magnon mode that can be 
considered as a pair of confined spinons. Only the reciprocal-space points 
$(\pi,0)$ and $(0,\pi)$ appear resistant to confinement, with their spectrum 
retaining a continuum form above its lower edge and their spinon-pair 
excitations remaining distributed across the entire system. 

Our results reinforce the deduction \cite{DallaPiazza} that deconfinement in 
a 2D system is a reciprocal-space phenomenon, and is not directly connected 
to the magnon energy. The fact that only two deconfined points survive on 
the square lattice, taken together with experimental \cite{Wessler2020} and 
numerical \cite{Gu2022,Hernandez2024} evidence that deconfinement is more 
widespread around the zone boundary in the $S = 1/2$ Heisenberg model on 
the honeycomb lattice, indicates that spinon confinement depends sensitively 
on the kinematics. The confinement transition appears as a smooth crossover, 
in that the suppression of spinonic behavior at all ``conventional'' 
${\bf k}$-vectors is a gradual phenomenon, not an abrupt one. Although 
the $(\pi,\pi)$ point is a weakness of the staggered-flux wave function, 
studies of variational states in 2D find long-range order to be a minor 
modification to a state of otherwise robust quantum fluctuations 
\cite{Liang1988}, and hence one expects that such modifications will 
have little effect on the magnetic excitation spectrum over most of the 
Brillouin zone. We remark that, because our studies consider finite systems, 
they do not exclude that spinon confinement occurs in the thermodynamic 
limit at all finite $\gamma$ at all wave vectors, other than $(\pi,0)$ and 
$(0,\pi)$, but they do show that there is no evidence for a confinement 
transition at some finite $\gamma$. Finally, it is important to remember 
that both spinons and magnons are quasiparticle approximations to the 
true many-body excited states, and as such both pictures capture different 
aspects of these.

{\it Acknowledgments.--}We thank the European Research Council for 
financial support through the Synergy network HERO (Grant No.~810451) 
and the Swiss National Science Foundation for support through Project Grant 
No.~188648.

\bigskip\bigskip\bigskip

\onecolumngrid

\setcounter{figure}{0}
\renewcommand{\thefigure}{S\arabic{figure}}

\setcounter{section}{0}
\renewcommand{\thesection}{S\arabic{section}}

\setcounter{equation}{0}
\renewcommand{\theequation}{S\arabic{equation}}

\bigskip\bigskip
\noindent
{\large{\bf Supplemental Materials to accompany the manuscript}}
\bigskip

\noindent
{\large{\bf Confined and deconfined spinon excitations in the rectangular-lattice quantum antiferromagnet}}
\bigskip

\noindent
N. E. Shaik, E. Fogh, B. Dalla Piazza, B. Normand, D. Ivanov, and H. M. R{\o}nnow
\bigskip\bigskip

\twocolumngrid

\section{S1.~~~~Spin excitations in the staggered-flux state}

\subsection{A.~~~~Staggered-flux wave function}

In order to provide a self-contained presentation, here we summarize the Gutzwiller-projected variational wave-function approach using the staggered-flux Ansatz. The construction of the wave function and optimization of the variational parameters are described in detail in Ref.~\cite{Shaik2020}.

The Heisenberg interaction may be expressed in terms of auxiliary (``slave'') fermions as 
\begin{eqnarray} \label{fermionHamil}
\mathcal{H}_{\langle i,j \rangle} & = & J_{ij} {\bf S}_i \! \cdot \! {\bf S}_j \\
& = & - {\textstyle \frac12} J_{ij} \big[ n_i \big( {\textstyle \frac12} n_j - 1 \big) + \sum_{\alpha\beta} c_{i\alpha}^\dagger c_{j\alpha} c_{j\beta}^\dagger c_{i\beta} \big], \nonumber 
\end{eqnarray}
where $i,j$ are pairs of nearest-neighbor sites, $\alpha$ and $\beta$ are spin flavors, and $n_i$ is the fermion number operator at site $i$. The first term is constant at half electronic filling and is neglected henceforth. The second term may be treated at the mean-field level by a standard decoupling scheme, and to describe the spin excitations in the absence of magnetic order \cite{DallaPiazza} we consider the staggered-flux fermionic
Hamiltonian
\begin{equation}
H_{\rm SF} = - \sum_{\langle i,j \rangle, \sigma} \chi_{i,j} c^\dagger_{i\sigma} c^{\phantom{\dagger}}_{j\sigma}
\label{H-mean-field}
\end{equation}
with the hopping amplitudes
\begin{eqnarray}
\chi_{i,i+x} & = & \chi_x e^{i(-1)^{i_x+i_y} \varphi/4}\, ,\\
\chi_{i,i+y} & = & \chi_y e^{-i(-1)^{i_x+i_y} \varphi/4}.
\end{eqnarray}
The ground state of $H_{\rm SF}$ depends on the two parameters $\alpha = \chi_x/\chi_y$ and $\varphi$, whose systematic dependence on $\gamma$ was studied in Ref.~\cite{Shaik2020}. 

Diagonalizing $H_{\rm SF}$ \eqref{H-mean-field} yields the eigenenergies 
\begin{equation}\label{Delta}
\varepsilon^\pm_{\bf k} = \pm {\textstyle \frac12} \big| \chi_x e^{i\varphi/4}\cos k_x + \chi_y \ e^{-i\varphi/4} \cos k_y \big|,
\end{equation}
which describe the dispersion relation of spinon excitations that are gapless at point nodes located at $(\pm \pi/2,\pm \pi/2)$. The corresponding eigenoperators are $d^\pm_{{\bf k} \sigma}$ and $d^{\pm \dagger}_{{\bf k} \sigma}$, which are combinations of the $c$ and $c^\dag$ operators that respectively annihilate or create spinons of spin $\sigma$ in a lower ($-$) or an upper ($+$) band in the magnetic Brillouin zone. The ground state of $H_{\rm SF}$ may then be expressed as 
\begin{equation}
\big| \psi_{\rm SF} \big\rangle = \prod_{{\bf k} \in {\rm MBZ}} d_{{\bf k}\uparrow}^{-\dagger} d_{{\bf k}\downarrow}^{-\dagger} | 0 \rangle, 
\end{equation}
where $| 0 \rangle$ is the fermionic vacuum. This wave function spans an eigenspace that includes sites in real space with double fermion occupation. The physical space of states with exactly one fermion per site is reached by applying a Gutzwiller projector, $P_{\rm G}$, to obtain the ground-state wave function
\begin{equation}
\big| \psi_{\rm GS} \big\rangle = P_{\rm G} \big| \psi_{\rm SF} \big\rangle.
\end{equation}

\subsection{B.~~~~Spin Excitations}
 
We construct spin excitations from the variational wave function as Gutzwiller-projected pairs of spinons (fermions and holes). Here we summarize the procedure described in Ref.~\cite{DallaPiazza}. We refer to spin-flipping excitations ($\sigma \leftrightarrow {\bar \sigma}$) as transverse and to non-spin-flip processes as longitudinal; because we consider the staggered-flux state without building in magnetic order, the transverse ($S^+_{\bf q}$) and longitudinal ($S^z_{\bf q}$) components should be equivalent and we focus only on the former. The basis vectors of the excitation subspace are given by applying the Gutzwiller projector to the space of mean-field states with a single particle-hole pair,
\begin{equation}
| {\bf k}, {\bf q} \rangle = P_{\rm G} \, d_{{\bf k}\uparrow}^{+\dagger} d^-_{{\bf k}-{\bf q}\downarrow} \big| \psi_{\rm SF} \big\rangle
\end{equation}
where ${\bf q}$ the physical momentum of the spin excitation and ${\bf k}$ is the internal relative momentum of the spinons. The eigenstates in this excitation subspace may then be expressed in the form
\begin{equation}
\big| \psi^n_{\bf q} \big\rangle = \sum_{\bf k} \phi_{\bf {kq}}^n | {\bf k}, {\bf q} \rangle, 
\label{eigenstates}
\end{equation}
where $n$ labels the number of the state in ascending order of energy. Diagonalization of the Hamiltonian [Eq.~(1) of the main text] in the non-orthogonal $| {\bf k}, {\bf q} \rangle$ basis is performed by solving the generalized eigenvalue problem
\begin{equation}
\sum_{{\bf k}'} H_{{\bf k}{\bf k}'}^q \phi_{{\bf k}'{\bf q}}^n =
E_{\bf q}^n
\sum_{\bf {k}'} O_{{\bf k}{\bf k}'}^q \phi_{\bf {k}'{\bf q}}^n \, ,
\label{generalized-eigenvalue}
\end{equation}
where
\begin{equation}
H_{{\bf k}{\bf k}'}^q = \langle {\bf k}, {\bf q} | \mathcal{H} | {\bf k}', {\bf q} \rangle, \;\;
O_{{\bf k}{\bf k}'}^q = \langle {\bf k}, {\bf q} | {\bf k}', {\bf q} \rangle,
\end{equation}
and we assume the eigenstates of Eq.~(\ref{eigenstates}) to be correctly normalized.

To quantify the real-space structure of the spin excitations, we define the state 
\begin{equation}
\big| \psi_{\bf q}^+ ({\bf r}) \big\rangle = P_{\rm G} \sum_{{\bf r}'} e^{i {\bf q} \cdot {\bf r}'} d^{+\dag}_{{\bf r} + {\bf r}' \uparrow} d^-_{{\bf r}' \downarrow} \big| \psi_{\rm SF} \big\rangle, 
\end{equation}
which describes a spin-flip operation for a pair of spinons with propagation vector ${\bf q}$ that are separated by a distance ${\bf r}$. In the momentum basis, this delocalized spin flip can be expressed as 
\begin{equation}
\big| \psi_{\bf q}^+ ({\bf r}) \big\rangle = \sum_{\bf k} \phi^+_{\bf {kq}} ({\bf r}) |{\bf k}, {\bf q} \rangle,  
\end{equation}
where the coefficients $\phi^+_{\bf {kq}} ({\bf r})$ can be evaluated from the coefficients of the Bogoliubov transformation defining $d^\pm_{{\bf k}\sigma}$ and $d^{\pm\dag}_{{\bf k}\sigma}$ \cite{DallaPiazza}. The quantity we use in Eq.~(4) of the main text to characterize the spinon separation can be understood as a time-averaged density of spinons prepared initially at zero separation: taking 
\begin{equation}
\rho^q ({\bf r},t) = \frac
{\big| \big\langle \psi^+_{\bf q} ({\bf r}) \big| e^{-i {\cal H} t} \big| \psi^+_{\bf q} ({\bf 0}) \big\rangle \big|^2}
{\big\langle \psi^+_{\bf q} ({\bf r}) \big| \psi^+_{\bf q} ({\bf r}) \big\rangle},
\label{rho-definition}
\end{equation}
introducing $\sum_n \big| \psi^n_{\bf q} \big\rangle \big\langle \psi^n_{\bf q} \big|$ as a resolution of the identity, and integrating the $e^{-i E_{\bf q}^n t}$ factors over all time leads to 
\begin{multline}
\rho^q ({\bf r})  = 
\lim_{T \rightarrow \infty} \frac{1}{T} \int_0^T \rho^q ({\bf r},t) \; dt  \\
= \sum_n \frac{ 
| \langle \psi^+_{\bf q} ({\bf r}) | \psi^n_{\bf q} \rangle
  \langle \psi^n_{\bf q} | \psi^+_{\bf q} ({\bf 0}) \rangle |^2}
{ \langle \psi^+_{\bf q} ({\bf r}) | \psi^+_{\bf q} ({\bf r}) \rangle}.
\end{multline}
We note here that our definition [Eq.~\eqref{rho-definition}] differs slightly from that of Ref.~\cite{DallaPiazza}, in that we have chosen to normalize the states $| \psi^+_{\bf q} ({\bf r}) \rangle$ to provide a more transparent interpretation of this measure without changing the qualitative results.

\subsection{C.~~~~Monte Carlo calculations}

Following Ref.~\cite{DallaPiazza}, the Gutzwiller projector in the excitation subspace can be written in the form $P_{\rm G} = \sum_\alpha | \alpha \rangle \langle \alpha |$, where $\{ | \alpha \rangle \}$ is the set of states with a single transverse (particle-hole) excitation that is also singly occupied in the site basis, meaning that for a system of $N = L$$\times$$L$ sites it has ${\textstyle \frac12} N + 1$ up and ${\textstyle \frac12} N - 1$ down spins. Because this space grows exponentially with $N$, the Hamiltonian in the particle-hole space can only be computed for a small system, or for a larger finite system by an approximate method, for which we adopt an approach of Monte Carlo sampling. We rearrange the elements of the Hamiltonian in the form
\begin{eqnarray}
\frac{H_{{\bf k},{\bf k}'}^q}{W_{\bf q}} & = & \sum_\alpha \frac{W_{\bf q} (\alpha)}{W_{\bf q}} \sum_\beta \frac{\langle {\bf k}, {\bf q} | \alpha \rangle \langle \alpha | \mathcal{H} | \beta \rangle \langle \beta | {\bf k}', {\bf q} \rangle}{W_{\bf q} (\alpha)} \nonumber \\ & \equiv & \sum_\alpha \rho(\alpha) f(\alpha), \nonumber
\end{eqnarray}
where 
\begin{eqnarray}
W_{\bf q} = \sum_\alpha W_{\bf q} (\alpha) = \sum_{{\bf k} \alpha} | \langle \alpha | {\bf k}, {\bf q} \rangle |^2 \\
= \sum_{\bf k} \langle {\bf k}, {\bf q} | {\bf k}, {\bf q} \rangle = {\rm Tr} \; O^q
\end{eqnarray}
expresses the normalization arising due to the non-orthonormal nature of the $| {\bf k}, {\bf q} \rangle$ basis. Because the elements of the Hamiltonian matrix appear now as a weighted average over a function, schematically $f(\alpha)$, with weights $\rho(\alpha)$, they can be estimated from Monte Carlo simulations performed in the space of states $\{ | \alpha \rangle\}$. 

\begin{figure*}[t]
\includegraphics[width=0.84\textwidth]{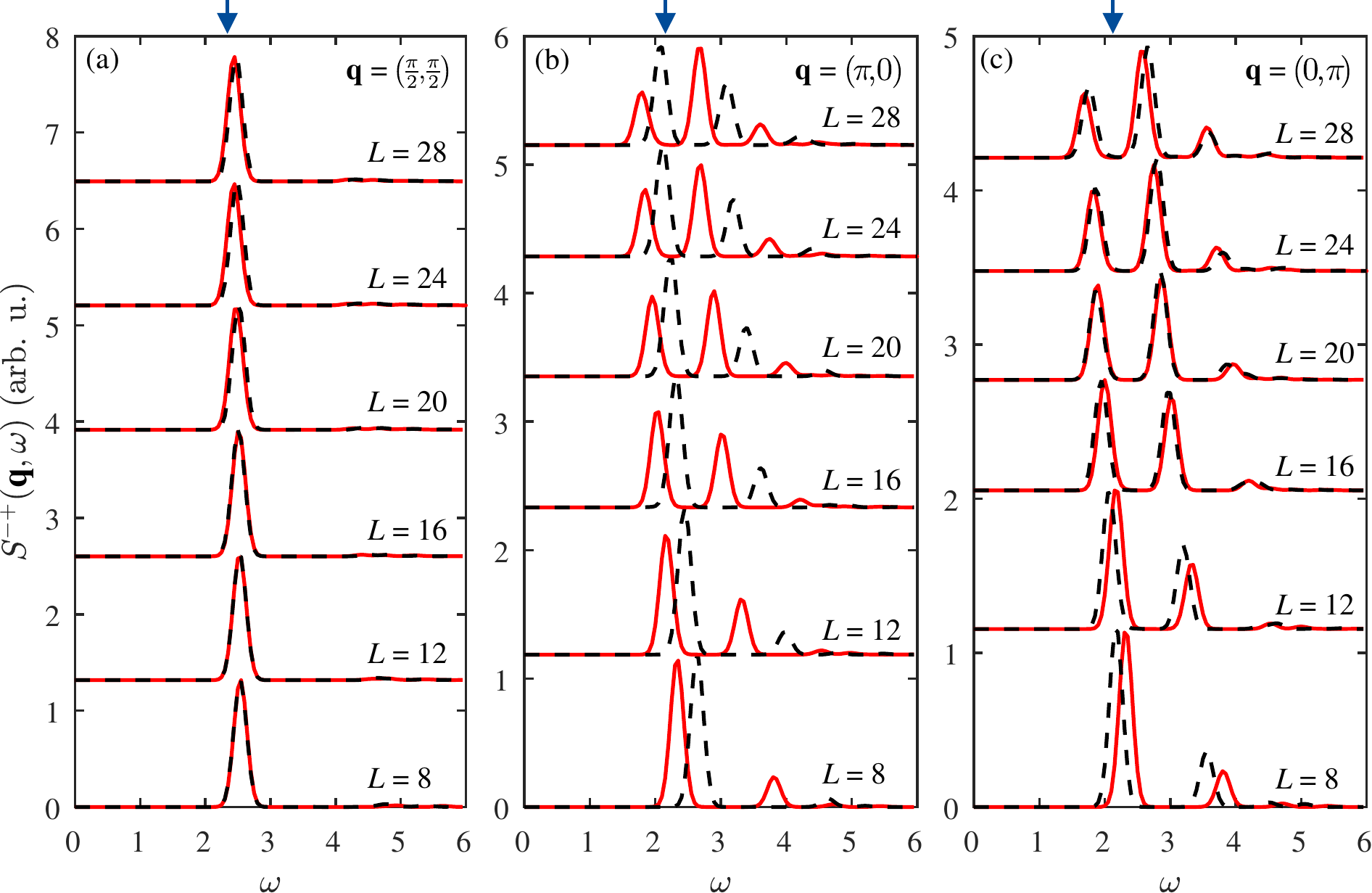}
\caption{{\bf Finite-size evolution of the TDSF.} TDSF shown for $\gamma = 1$ 
at $(\pi/2,\pi/2)$ (a), $(\pi,0)$ (b), and $(0,\pi)$ (c) for increasing system 
sizes, $L$. The solid red line shows results computed with ABCs in $x$ and the 
dashed black line PBCs. The arrows mark the lowest excitation energies expected 
from a combination of theoretical \cite{Manousakis}, numerical \cite{Shao2017,
Powalski2018}, and experimental \cite{Plumb2014,DallaPiazza} methods, which 
benchmark the rate of convergence of our results with system size.} 
\label{fig:figs1}
\end{figure*}

In a similar way, we compute the overlap matrix $O_{{\bf k}{\bf k}'}^q / W_{\bf q}$. The overall normalization of $H_{{\bf k}{\bf k}'}^q$ and $O_{{\bf k}{\bf k}'}^q$ by $W_{\bf q}$ is not relevant for the generalized eigenvalue problem of Eq.~\eqref{generalized-eigenvalue}, but is important to take into account for the calculation of other quantities. An important example for our purposes is the TDSF, which we calculate as
\begin{multline}
\label{etdsf}
S^\pm ({\bf q},\omega) = \sum_n
\big\langle \psi_{\rm GS} | S_{\bf q}^- | \psi_{\bf q}^n \big\rangle
\big\langle \psi_{\bf q}^n | S_{\bf q}^+ | \psi_{\rm GS} \big\rangle \,
\delta (\omega - \omega_n) \\
 =  \sum_{n}
\Big| \sum_{\bf {kk}'} \phi_{\bf {kq}}^{n*} O_{\bf {kk}'}^q \phi_{{\bf k}'{\bf q}}^+ ({\bf 0}) \Big|^2 \, 
\delta (\omega - \omega_n).
\end{multline}
Here the normalization at each ${\bf q}$ is lost because of factors of $W_{\bf q}$, and can be restored by using the sum rule equating the energy-integrated TDSF to the instantaneous transverse spin correlation function \cite{DallaPiazza}.

\section{S2.~~~~Finite-size analysis of the TDSF at selected high-symmetry points}

All of our calculations are performed on square systems ($L_x = L_y = L$) with 
even $L$, maintaining ABCs in the $y$ direction, while in the $x$ direction we 
compare both PBCs and ABCs. For practical purposes, in addition to avoiding the 
nodal points [$k = (\pm \pi/2, \pm \pi/2)$], the comparison highlights that 
modes well defined in ${\bf k}$-space are largely independent of the BCs, 
whereas peaks appearing as the finite-size analog of a continuum change their 
positions significantly when the BCs are changed. 

This physics offers direct numerical insight into the continuum or discrete 
nature of the spectrum as the thermodynamic limit is approached. In 
Fig.~\ref{fig:figs1} we show the evolution of the TDSF at $\gamma = 1$ 
with the system size for both types of $x$-axis BCs. Considering first 
the size effect, increasing $L$ causes very little change at the magnon-like 
$(\pi/2,\pi/2)$ point, whereas at $(\pi,0)$ and $(0,\pi)$ we find an increasing 
number of equally spaced modes, suggesting a continuum as $L \rightarrow 
\infty$. The loss of spectral weight in the lowest-energy mode at $(\pi,0)$ 
with ABCs [Fig.~\ref{fig:figs1}(b)], and the gain in the second mode, confirms 
the statement made in the main text about the correct identification of the 
magnon branch (which is the lower bound of the emerging continuum). 

Turning to the effect of the BCs, this is clearest at large $L$, where the 
single peak at $(\pi/2,\pi/2)$ is barely affected [Fig.~\ref{fig:figs1}(a)], 
whereas the multiple peaks at $(\pi,0)$ shift by approximately half of their 
energetic spacing [Fig.~\ref{fig:figs1}(b)]. At $(0,\pi)$ the peak positions 
are little affected by the $x$-axis BCs [Fig.~\ref{fig:figs1}(c)], but would 
change with the $y$-axis BCs as in Fig.~\ref{fig:figs1}(b).

\section{S3.~~~~Self-consistent spin-wave theory}

To gain further insight into the spin excitation spectra of the rectangular 
lattice with different $\gamma$, we also performed calculations of the 
one-magnon excitation branch expected within spin-wave theory. For an accurate 
analysis beyond the linear level, we follow the self-consistent method 
introduced by Oguchi \cite{Oguchi} to compute the spin-wave spectrum at 
fourth order in the Holstein-Primakoff \cite{rhp} operators $a_i^\dag$ and 
$a_i$, which are respectively bosonic creation and annihilation operators 
for a $\Delta S = 1$ process on lattice site $i$. This level of treatment is 
equivalent to order $1/S^2$, or the Hartree-Fock diagrams.

Applying the Wick theorem to the four-operator terms requires the definition 
of the expectation values, 
\begin{eqnarray}
n & = & \langle a_i^\dag a_i \rangle, \qquad\qquad\qquad \; t_\tau = \langle 
a_i^\dag a_{i+\tau} \rangle = \langle a_i a_{i+\tau}^\dag \rangle,\nonumber \\
\delta & = & \langle a_i a_i \rangle = \langle a_i^\dag a_i^\dag \rangle, 
\qquad \Delta_\tau = \langle a_i a_{i+\tau} \rangle = \langle a_i^\dag 
a_{i+\tau}^\dag \rangle, \nonumber
\end{eqnarray}
where we restrict $\tau \in {\hat{\bf x}}, {\hat{\bf y}}$ to nearest-neighbor sites 
only. These quantities are taken to be uniform for all sites of the rectangular 
lattice, although $t_\tau$ and $\Delta_\tau$ can differ with the direction of 
$\tau$. By treating these expectation values as mean fields, the quartic boson 
terms are decoupled to a quadratic form and the spin-wave dispersion relation 
is altered from the conventional linear spin-wave form for an antiferromagnet, 
\begin{equation}
\omega^0_k = \sqrt{A_k^2 + B_k^2}
\end{equation}
to the self-consistent form 
\begin{equation}
\label{escsw}
\omega_k = \sqrt{(A_k + dA_k)^2 + (B_k + dB_k)^2}.
\end{equation}

\begin{figure}[t]
\includegraphics[width=0.95\columnwidth]{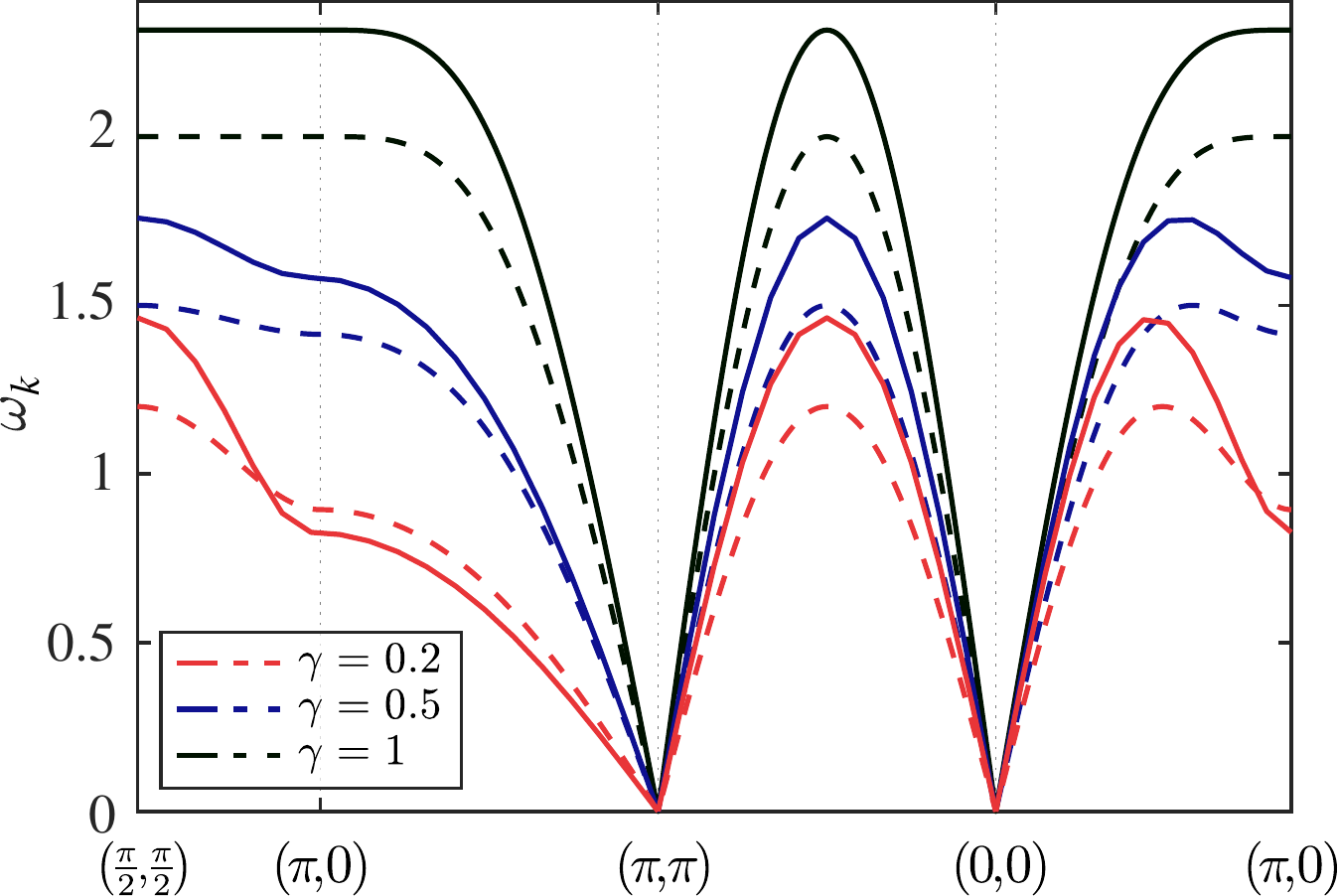}
\caption{{\bf Self-consistent spin-wave spectra.} Renormalized spin-wave 
dispersion, $\omega_k$, calculated from Eq.~\eqref{escsw} (solid lines) 
and compared with the magnon dispersion of linear spin-wave theory (dashed 
lines) for $\gamma = 0.2$, 0.5, and 1.}
\label{fig:figs2}
\end{figure}

The mean fields can be computed from the ground state of the quadratic 
Hamiltonian in terms of the standard (hyperbolic Bogoliubov) coefficients 
\begin{equation}
u_k = \sqrt{\frac12 \left( \frac{A_k}{\omega_k^0} + 1 \right)}, \quad
v_k = {\rm sgn}(B_k) \sqrt{\frac12 \left( \frac{A_k}{\omega_k^0} - 1 \right)}, \nonumber
\end{equation}
as 
\begin{eqnarray}
n & = & \frac{1}{N} \sum_k v_k^2, \qquad \;\;\;\;\; t_\tau = \frac{1}{N} \sum_k \cos (k \cdot \tau) v_k^2, \nonumber \\
\delta & = & \frac{1}{N} \sum_k u_k v_k, \qquad \Delta_\tau = \frac{1}{N} \sum_k \cos (k \cdot \tau) u_k v_k. \nonumber
\end{eqnarray}
Symmetry ensures that $u_{k \pm Q} = u_k$ and $v_{k \pm Q}
 = - v_k$, as a result of which $\delta = 0$. Similarly, with $\tau$ including 
only nearest-neighbor sites, $t_\tau$ also vanishes. The renormalization terms 
in the self-consistent spin-wave dispersion [Eq.~\eqref{escsw}] then become 
\begin{equation}
dA_k = \sum_\tau J_\tau dA_k^\tau \;\; {\rm and} \;\; dB_k = \sum_\tau J_\tau dB_k^\tau 
\end{equation}
with 
\begin{equation}
dA_k^\tau = 2(\Delta_\tau - n) \;\; {\rm and} \;\; dB_k^\tau = 2 (n - \Delta_\tau) \cos (k \cdot \tau).
\end{equation}

The renormalization of the linear spin-wave energies can be codified in the 
factor $Z_c({\bf k}) = \omega_k/\omega^0_k$. At $\gamma = 1$, $Z_c
 = 1.158$ becomes {\bf k}-independent, providing an excellent approximation 
to the correction $Z_c = 1.18$ deduced from the nonlinear sigma model 
\cite{Manousakis}. Figure \ref{fig:figs2} shows the spin-wave dispersions 
computed from Eq.~\eqref{escsw} for three different values of $\gamma$, which 
illustrate the qualitative extent both of the correction and of the 
{\bf k}-dependence at arbitrary $\gamma$. The dispersion relations for 
$\gamma = 0.5$ and 1 in Fig.~\ref{fig:figs2} are shown as the black lines 
in Figs.~2(c) and 2(d) of the main text.

\end{document}